\documentclass[usenatbib]{emulateapj}
\submitted{{\it Accepted for publication in ApJ}}
\shortauthors{Cowan et al.}
\shorttitle{Alien Maps}
\begin{document}

\title{Alien Maps of an Ocean-Bearing World}

\author{Nicolas B. Cowan\altaffilmark{1}, Eric Agol, Victoria S. Meadows\altaffilmark{2}, Tyler Robinson\altaffilmark{2},} 
\affil{Astronomy Department and Astrobiology Program, \\University of Washington, Box 351580, Seattle, WA  98195}
\author{Timothy A. Livengood\altaffilmark{3}, Drake Deming\altaffilmark{2},}
\affil{NASA Goddard Space Flight Center, Greenbelt, MD 20771}
\author{Carey M. Lisse,}
\affil{Johns Hopkins University Applied Physics Laboratory, SD/SRE, MP3-E167, \\11100 Johns Hopkins Road, Laurel, MD 20723}
\author{Michael F. A'Hearn, Dennis D. Wellnitz,}
\affil{Department of Astronomy, University of Maryland, College Park MD 20742}
\author{Sara Seager,}
\affil{Department of Earth, Atmospheric, and Planetary Sciences, Dept of Physics, Massachusetts Institute of Technology, 77 Massachusetts Ave. 54-1626, MA 02139}
\author{David Charbonneau,}
\affil{Harvard-Smithsonian Center for Astrophysics, 60 Garden Street, Cambridge, MA 02138}
\author{and the EPOXI Team}
\altaffiltext{1}{Email: cowan@astro.washington.edu}
\altaffiltext{2}{NASA Astrobiology Institute Member}
\altaffiltext{3}{Also with the Department of Astronomy, University of Maryland}

\begin{abstract}
When Earth-mass extrasolar planets first become detectable, one challenge will be to determine which of these worlds harbor liquid water, a widely used criterion for habitability. Some of the first observations of these planets will consist of disc-averaged, time-resolved broadband photometry. To simulate such data, the Deep Impact spacecraft obtained light curves of Earth at seven wavebands spanning 300--1000~nm as part of the EPOXI mission of opportunity. In this paper we analyze disc-integrated light curves, treating Earth as if it were an exoplanet, to determine if we can detect the presence of oceans and continents. We present two observations each spanning one day, taken at gibbous phases of $57^{\circ}$ and $77^{\circ}$, respectively. As expected, the time-averaged spectrum of Earth is blue at short wavelengths due to Rayleigh scattering, and gray redward of 600 nm due to reflective clouds. The rotation of the planet leads to diurnal albedo variations of 15--30\%, with the largest relative changes occuring at the reddest wavelengths. To characterize these variations in an unbiased manner we carry out a principal component analysis of the multi-band light curves;  this analysis reveals that 98\% of the diurnal color changes of Earth are due to only 2 dominant eigencolors. We use the time-variations of these two eigencolors to construct longitudinal maps of the Earth, treating it as a non-uniform Lambert sphere. We find that the spectral and spatial distributions of the eigencolors correspond to cloud-free continents and oceans; this despite the fact that our observations were taken on days with typical cloud cover. We also find that the near-infrared wavebands are particularly useful in distinguishing between land and water. Based on this experiment we conclude that it should be possible to infer the existence of water oceans on exoplanets with time-resolved broadband observations taken by a large space-based coronagraphic telescope.
\end{abstract}


\section{Introduction}
The rate of discovery of extrasolar planets is increasing and every
year it is possible to detect smaller planets. It is only a
matter of time before we detect Earth-analogs, but even then our
ability to study them will remain limited. Due to exoplanets'
great distance from us and their relative faintness, spatially
resolving them from their host stars is only currently possible for
hot, young Jovian planets in long-period orbits \citep{Kalas_2008,
Marois_2008, Lagrange_2008}. For other planets ---including Solar System analogs---
spatially separating the image of the planet from that of the host
star will have to wait for space-based telescopes like TPF/Darwin
\citep{Traub_2006, Beichman_2006, Fridlund_2002}. But even these
telescopes will not have sufficient angular resolution to spatially resolve
the disc of an exoplanet.

As noted over a century ago by \cite{Russell_1906}, variations in the reflected light of an
unresolved rotating object can be used to learn about albedo markings
on the body. Such light curve inversions have
proved valuable to interpret the photometry of objects viewed near
full phase and led, for example, to the first albedo map of Pluto
\citep{Lacis_1972}. More recently, thermal light curves have made it possible to measure the day/night temperature contrast of short-period exoplanets \citep{Harrington_2006, Cowan_2007, Snellen_2009}. Light curve inversion \citep{Cowan_2008} has even been used to construct coarse longitudinal thermal maps of hot Jupiters \citep{Knutson_2007, Knutson_2009}.

The optical and near-IR light curves of Earth, on the other hand, have not been thoroughly studied to date. Earthshine, the faint illumination of the dark side of the Moon due to reflected light from Earth, has been used to study the reflectance spectrum, cloud cover variability, vegetation signatures and the effects of specular reflection for limited regions of our planet \citep{Goode_2001, Woolf_2002, Qiu_2003, Palle_2003, Palle_2004, Montanes_2005, Seager_2005, Hamdani_2006, Montanes_2006, Montanes_2007, Langford_2009}. Brief snapshots of Earth obtained with the Galileo spacecraft have been used to study our planet \citep{Sagan_1993, Geissler_1995} and numerical models have been developed to predict how diurnal variations in disc-integrated light could be used to characterize Earth \citep{Ford_2001, Tinetti_2006, Tinetti_2006b, Palle_2008, Williams_2008}. 

This paper is the first in a series analyzing the photometry and spectroscopy of Earth obtained as part of the EPOXI mission and is written in a different spirit than most studies of Earthshine. Rather than attempting to produce a detailed model which exactly fits the observations (a.k.a. ``forward modeling''), we make a few reasonable simplifying assumptions which allow us to extract information directly from the data (``backward modeling''). This approach is complementary to detailed modeling and will be especially appropriate when studying an alien world with limited data. This paper is organized as follows: in \S~2 we describe the time-resolved observations of the entire disc of Earth used in this study; in \S~3 we use principal component analysis to determine the dominant spectral components of the planet in a model-independent way; we use light curve inversion in \S~4 to convert the diurnal albedo variations into a longitudinal map of Earth; we discuss our results in \S~5; our conclusions are in \S~6.

\section{Observations}
The EPOXI\footnote{The University of Maryland leads the overall EPOXI mission, including the flyby of comet Hartley 2. NASA Goddard leads the exoplanet and Earth observations.} mission reuses the still-functioning Deep Impact spacecraft
that successfully observed comet 9P/Tempel~1. EPOXI science targets
include several transiting exoplanets and Earth \emph{en route} to a
flyby of comet 103P/Hartley~2. The EPOXI Earth observations are valuable
for exoplanet studies because they are the first time-resolved,
multi-waveband observations of the full disc of Earth. These
data reveal Earth as it would appear to observers on an extrasolar planet, and can only be obtained from a relatively distant vantage
point, not from low-Earth orbit.  The data consist of an equinox observation on March 18, and near solstice on
June 6. An observation taken on May 28 included a planned lunar
transit, but we save the analysis of those data for a future paper. The observations, summarized in Table~\ref{observations}, were taken when Earth was in a partially illuminated ---gibbous--- phase. With the rare exception of transiting planets, this is the phase at which habitable exoplanets will be observed.

\begin{table*}[htb]
\begin{minipage}{126mm}
\caption{\bf EPOXI Earth observing campaigns}
\label{observations}
\begin{tabular}{@{}lcccccc}
\hline
Date & Starting & Phase & Illuminated Fraction & Spacecraft & Angular Diam. & Pixels Spanned\\
 & CML &  & of Earth Disc & Range & of Earth & by Earth\\ 
\hline
3/18/2008 & $150^{\circ}$ W & 57$^{\circ}$ & 77\% &  0.18 AU & 1.63$^{\prime}$ & $\sim 240$\\
5/28/2008 & $195^{\circ}$ W & 75$^{\circ}$ & 63\% & 0.33 AU & 0.89$^{\prime}$ & $\sim 130*$\\
6/4/2008 & $150^{\circ}$ W & 77$^{\circ}$ & 62\% & 0.34 AU & 0.87$^{\prime}$ & $\sim 130$\\
 \hline
\end{tabular}

\medskip
The CML is the Central Meridian Longitude, the longitude of the sub-observer point. The planetary phase, $\alpha$, is the star--planet--observer angle and is related to the illuminated fraction by $f = \frac{1}{2}(1+\cos\alpha)$. *The 5/28/2008 observation is not used in this paper due to a planned lunar transit.
\end{minipage}
\end{table*}

\begin{figure}[htb]
\includegraphics[width=84mm]{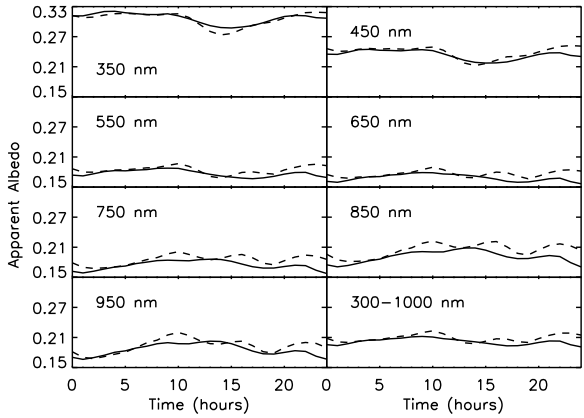}
\caption{Seven light curves obtained by the EPOXI spacecraft on March 18 (solid lines) and on June 4, 2008 (dashed lines). The bottom-right panel shows changes in the bolometric albedo of Earth.}
\label{combined_lightcurves}
\end{figure}

Deep Impact's 30~cm diameter telescope coupled with the High Resolution Imager \citep[HRI,][]{Hampton_2005} recorded images of Earth in seven 100~nm wide optical
wavebands spanning 300--1000~nm. Hourly
observations were taken with the filters centered on 350, 750 and
950~nm, whereas the 450, 550, 650 and 850~nm data were taken every 15
minutes; each set of observations lasted 24 hrs. The exposure times for the different wavebands are: 73.4~msec at 350~nm; 13.3~msec at 450~nm; 8.5~msec at 550~nm; 9.5~msec at 650~nm; 13.5~msec at 750~nm; 26.5~msec at 850~nm; 61.5~msec at 950~nm. Although the EPOXI images of Earth offer spatial resolution of better than 100~km, we mimic the data that will eventually be
available for exoplanets by integrating the flux over the entire disc
of Earth and using only the hourly EPOXI observations from each of the wavebands, producing seven light curves for each of the two observing
campaigns, shown in
Figure~\ref{combined_lightcurves}. Our results are the same when we use the 450, 550, 650 and 850~nm data from :00, :15, :30 or :45. The photometric uncertainty in these data is exceedingly small: on the order of 0.1\% relative errors. 

Since we are interested in the properties of the planet rather than
its host star, we normalize the light curves by the average
solar flux in each bandpass using the solar
spectrum\footnote{We use the ASTM-E-490 Standard Solar Constant and Zero Air Mass Solar
Spectral Irradiance Tables: http://www.astm.org/Standards/E490.htm} to
obtain the reflectivity in each waveband. We express the brightness of
the planet as an apparent albedo, the average albedo of regions on the planet that are both visible and illuminated during each observation\footnote{The ratio of the observed flux to
the expected flux for a planet of the same size and phase exhibiting diffuse reflection and with an albedo of unity everywhere on its surface. Mathematical details of this definition are in \S~\ref{diffuse}.}. The spacecraft was above the middle
of the Pacific Ocean at the start of both observing campaigns, so the
shapes of the light curves are similar for both epochs. The June light
curves vary more rapidly because a smaller fraction of the illuminated hemisphere of Earth was
visible (62\% rather than 77\%) and thus less of the planet was averaged together in any
given frame.

\subsection{Cloud Variability}
Clouds cover roughly half of Earth at any point in time
\citep{Palle_2008} and they dominate the disc-integrated albedo of the planet \citep[e.g.,][]{Tinetti_2006}. Mapping surface features could be problematic if
large scale cloud formations move or disperse on timescales shorter than a planetary rotation. 

The necessary condition for variable cloud patterns ---surface pressure
and temperature near the condensation point of water---is a likely
precondition for habitability, and hence may pose a problem for the
planets that interest us the most. Changeable cloud cover may indicate the presence of water vapor in a planet's atmosphere, but here we are concerned with the presence of liquid water on the planet's surface.

\begin{figure}[htb]
\includegraphics[width=84mm]{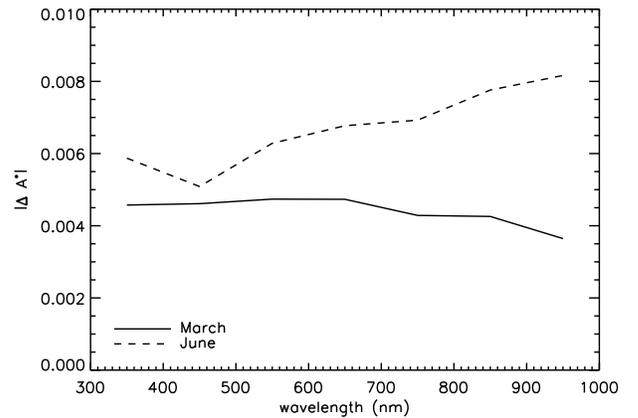}
\caption{The discrepancy in apparent albedo between the start and end of each observing campaign. Note that during the March observations the flux decreased at all seven wavebands, while in June the flux increased. We attribute these changes in albedo to changes in cloud cover.}
\label{cloud_variability}
\end{figure}

 After 24 hours of rotation the same hemisphere of Earth should be facing the Deep Impact Spacecraft so the integrated brightness of the planet's surface should be identical, provided one has accounted for the difference between the sidereal and solar day, as well as slight changes in the geocentric distance of the spacecraft and in the phase of the planet as seen from the spacecraft. Note that our observations are not taken near full phase so we may safely ignore the opposition effect \citep{Hapke_1993}. Even after correcting for all known geometric effects, the observed fluxes at the start and end of a given observing campaign differ by $2.2$\% and $3.4$\% for the March and June observing campaigns, respectively, as shown in Figure~\ref{cloud_variability}. We attribute this discrepancy to diurnal changes in cloud cover. Therefore, even though the EPOXI photometry is excellent, for the purposes of our model fits we use effective uncertainties equal to $|F_{\rm start} - F_{\rm end}|/2$ for each waveband.

Our 24-hour cloud variability of a few percent is somewhat smaller than estimates from Earthshine observations \citep[e.g.,][found day-to-day cloud variations of roughly 5\% and 10\%, respectively]{Goode_2001, Palle_2004}, and is a small effect compared to the rotational modulation of Earth's albedo or the 10--20\% differences between the light curves from the two EPOXI observing campaigns. The 10--30\% variations in apparent albedo we observe due to the Earth's rotation (with the largest variations occuring at near infrared wavebands) agrees with previous optical studies \citep[][find diurnal variations of 15--20\%]{Ford_2001, Goode_2001}.

The differences between the March and June observations could be due to some combination of stochastic changes in cloud cover, coherent (seasonal) changes in cloud cover, or simply a change in viewing geometry (which we discount in \S~4.3). Although daily changes in cloud cover are modest, the cloud cover will be entirely different for observations taken months apart \citep[see, for example, Figure~6 of][]{Palle_2008}.

\begin{figure}[htb]
\includegraphics[width=84mm]{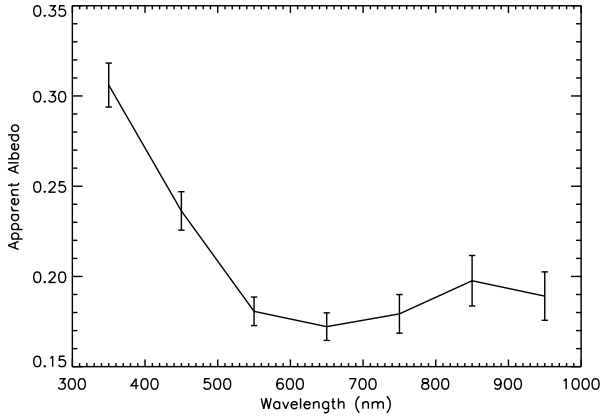}
\caption{Time-averaged broadband spectrum of Earth, based on EPOXI observations taken on 03/18/2008 and 06/04/2008 (the time-averaged spectra for the two epochs are indistinguishable). The error-bars show $1\sigma$ time-variability. The steep ramp at short wavelengths is due to Rayleigh scattering.  The near-IR wavebands exhibit the largest relative time-variability.}
\label{mean_spectrum}
\end{figure}

\section{Determining Principal Colors}
In this study we assume no prior knowledge of the different surface types of the unresolved planet. Our data consist of 50 broadband spectra of Earth (25 hourly observations for each of two epochs). The time-averaged spectrum of Earth is blue at short wavelengths due to Rayleigh scattering and gray longward of 550~nm because of clouds, as shown in Figure~\ref{mean_spectrum}. The changes in color of Earth during our observations can be thought of as occupying a 7-dimensional parameter space (one for each waveband). Principal component analysis \citep[PCA, e.g.][]{Connolly_1995} allows us to reduce the dimensionality of these data by defining orthogonal eigenvectors in the parameter space (eigencolors). Qualitatively, the observed spectrum of Earth at some time $t$ can be recovered using the equation:
\begin{equation} \label{pca}
A^{*}(t) = \langle A^{*}\rangle + \sum_{i=1}^{7} C_{i}(t) A_{i},
\end{equation}
where $\langle A^{*}\rangle$ is the time-averaged spectrum of Earth, $A_{i}$ are the seven orthogonal eigencolors, and $C_{i}$ are the instantaneous projections of Earth's colors on the eigencolors. The terms in the sum are ranked by the time-variance in $C_{i}$, from largest to smallest. Insofar as the projections are consistently small for the highest $i$, they can be ignored with only a minor penalty in goodness-of-fit.

\begin{figure}[htb]
\includegraphics[width=84mm]{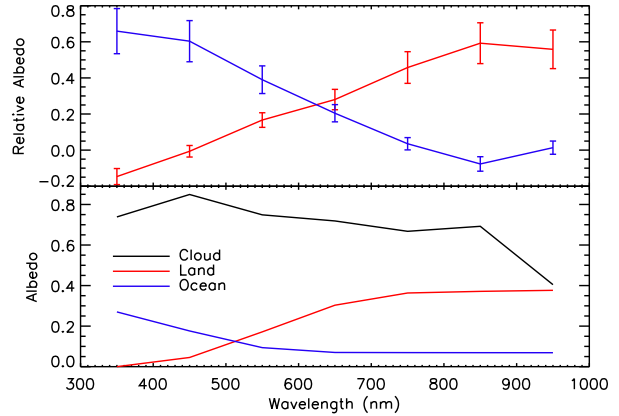}
\caption{Spectra for the two dominant eigencolors of Earth, as determined by PCA (top panel). For comparison, the bottom panel shows actual spectra of clouds, soil and oceans on Earth \citep{Tinetti_2006, McLinden_1997}.}
\label{eigenspectra}
\end{figure}

\begin{figure}[htb]
\includegraphics[width=84mm]{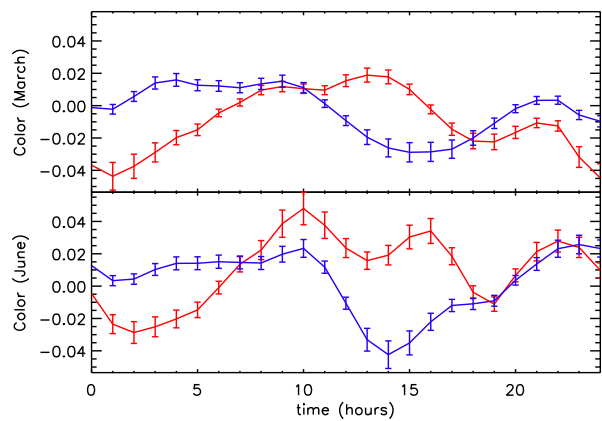}
\caption{Contributions of Earth's two principal colors, as determined by PCA, relative to the average Earth spectrum from \emph{both} epochs of observation. Each set of observations spans a full rotation of the planet, starting and ending with the spacecraft directly above $150^{\circ}$~W longitude, the middle of the Pacific Ocean.}
\label{eigenprojections}
\end{figure}

We find that 98\% of the changes in color do not occupy the whole parameter space but instead lie on a two-dimensional ``plane'' defined by the two principal eigencolors. That is to say, truncating the sum in Equation~\ref{pca} at $i=2$ only leads to errors of a couple percent. In detail, the plane has some thickness to it: two additional components only present at the 1--2$\sigma$ level that we neglect in the remainder of this study. We estimate the uncertainty in the PCA by creating 10,000 versions of the light curves with added Gaussian noise. The standard deviation in the resulting PCA parameters provides an estimate of their uncertainty. The primary eigencolors ($A_{1}$ and $A_{2}$ from Equation~\ref{pca}) are shown in Figure~\ref{eigenspectra} and their relative importance as a function of time ($C_{1}(t)$ and $C_{2}(t)$ from Equation~\ref{pca}) is shown in Figure~\ref{eigenprojections}. The uncertainties on the PCA are correlated but we represent them as error bars in the figures.

The eigencolors should not be thought of as spectra of different surface types. Rather, they are particular combinations of filters which are most sensitive to the different surface types rotating in and out of view. As such, the eigencolors are a \emph{relative} color from the Earth mean. The first eigencolor is most sensitive to variations in the red wavebands (since the albedo of Earth varies the most at near-IR wavelengths) and the second eigencolor is most sensitive to the blue wavebands. 

Since the mean Earth spectrum is ---to first order--- a cloud spectrum seen through a scattering atmosphere, the principal colors of Earth are related to the main surface types on Earth: cloud-free continents are most reflective at longer wavelengths \citep[][constructed from the ASTER Spectral Library found at http://speclib.jpl.nasa.gov]{Tinetti_2006}, while cloud-free oceans are most reflective in the blue \citep{McLinden_1997}. For example, the presence (or lack) of continents shows up as positive (or negative) excursions of the red eigencolor. The relative contributions of the colors are the projection of the Earth's instantaneous color onto an eigencolor. They are not identical in the March and June observations due to different cloud cover: when both the red and blue eigencolors are positive, there was more than average cloud cover, while regions with uniformly low eigencolors correspond to relatively cloud-free regions. Nevertheless, the similar shapes of the two sets of curves would indicate to extrasolar observers that the principal colors are most sensitive to fixed surface features that were visible from one season to the next (We run the PCA on both the March and June observations simultaneously.). The implicit ``third surface'' in this analysis is the time-averaged spectrum of the Earth, which includes clouds and Rayleigh scattering.

To test how sensitively the results of the PCA depend on photometric uncertainty, we repeat the analysis with additional noise. Our principal result ---the significant detection of red and blue eigencolors and their temporal variations--- are essentially unchanged for photometric uncertainties smaller than $2$--$3$\%. Observations of exoplanets of this quality are not around the corner, but can be obtained in the foreseeable future: a 16~m space telescope \citep[e.g.,][]{Postman_2009} equipped with a coronagraph could obtain 2\% photometry with 1 hour exposures of an Earth-analog at 10~pc. We have assumed a planet with the same radius and mean albedo as Earth, orbiting at 1~AU from a Sun-like star and observed near quadrature. We compute signal-to-noise as in \cite{Agol_2007} including photon counting noise from the planet and PSF noise from the host star, but neglecting zodiacal and exozodiacal noise.

\section{Mapping Surface Types}
The time-variation in the eigencolors, Figure ~\ref{eigenprojections}, tells us about the spatial variations of the colors around the planet. A region on the planet contributes more or less to the disc-integrated light depending on the amount of sunlight the region is receiving, its projected area as seen by the observer, and its albedo. As time passes, different regions of the planet rotate through the ``sweet-spot'' where the combination of illumination and visibility is optimal. The zeroth order approach to mapping the planet would be to assume that \emph{all} of the light from the planet originates from this region, as would be the case for purely specular reflection. The observed light curve could then be directly converted into a longitudinal albedo map of the planet at the appropriate latitude. We explore this limiting case in Appendix~III.    

More exactly, determining the spatial distribution of a color based on the planet's multi-band light curves is a deconvolution problem equivalent to mapping the albedo markings on a body based on its disc-integrated reflected light curve. This problem was solved by \cite{Russell_1906} for outer Solar System objects, which are always observed near full phase as seen from Earth. Exoplanets, however, appear at a variety of phases (E.g., crescent, gibbous) and therefore require a more complete solution. We keep the problem tractable by considering the diffusely reflecting (Lambertian) regime, in which surfaces reflect light equally in all directions.

Most materials are not perfectly Lambertian, instead 
scattering light preferentially backwards, forwards, or specularly. Coherent back-scattering is only significant near full phase \citep[the opposition effect,][]{Hapke_1993}. In fact, observations of Earthshine \citep{Qiu_2003, Palle_2003} and simulations \citep{Ford_2001, Williams_2008}, indicate that the
disc-integrated light of a cloudy planet like Earth is well described by diffuse reflection, provided the star--planet--observer angle is close to 90$^{\circ}$. Future missions ---interferometers,
coronagraphs, or occulters--- will ideally observe planets at phase angles slightly smaller than 90$^{\circ}$ (a.k.a. ``quadrature''), since that is when the S/N ratio is greatest \citep[E.g.,][]{Agol_2007}.

\subsection{Reflected Light From A Non-Uniform Lambert Sphere}\label{diffuse}
The flux from a diffusely reflective non-uniform sphere can be parametrized in terms of its albedo map, $A(\theta,\phi)$, where $\theta$ and $\phi$ are latitude and longitude on the planet, respectively. The visibility and illumination of a region on the planet at time $t$ are denoted by $V(\theta, \phi, t)$ and $I(\theta, \phi, t)$, respectively. $V$ is unity at the sub-observer point, drops as the cosine of the angle from the observer and is null on the far side of the planet from the observer; $I$ is unity at the sub-stellar point, drops as the cosine of the angle from the star and is null on the night-side of the planet. The mathematical details of $V$ and $I$ can be found in Appendix~I. The planet/star flux ratio, $\epsilon$, is obtained by integrating the product of visibility, illumination and albedo over the planet's surface: 
\begin{equation}
\epsilon = \frac{1}{\pi} \left( \frac{R_{p}}{a}\right)^{2} \oint V(\theta, \phi, t) I(\theta, \phi, t) A(\theta, \phi) d\Omega, 
\end{equation}
where $R_{p}$ is the planet's radius and $a$ is its mean orbital radius. The integral is over the \emph{entire} surface of the planet, but the integrand is only non-zero for the regions of the planet that are both visible and illuminated (e.g., for $\frac{1}{4}$ of a planet viewed at quadrature).

The flux ratio primarily depends on the planet's orbital phase, the observer--planet--star angle, and the ratio $(R_{p}/a)^2$. We define the apparent albedo, $A^{*}$, as the ratio of the flux from the planet divided by the flux we would expect \emph{at the same phase} for a perfectly reflecting ($A \equiv 1$) Lambert sphere \citep[see also][]{Qiu_2003}: 
\begin{equation}
A^{*}(t) = \frac{\oint V I A d\Omega}{\oint V I d\Omega}.
\end{equation}
A uniform planet would have an apparent albedo that is constant over a planetary rotation; a true Lambert sphere would further have a constant apparent albedo during the entire orbit. For non-transiting exoplanets, the planetary radius is unknown, so $A^{*}$ can only be determined to within a factor of $R_{p}^{2}$.

\subsection{Sinusoidal and N-Slice Maps}
The map $A(\theta, \phi)$ can take any form but diurnal brightness variations are due entirely to the longitude-dependence of albedo, provided that the planet's rotational period is much shorter than its orbital period. In the opposite extreme of a tidally-locked planet, seasonal variations in reflected light will be due to both albedo markings on the planet and changes in phase, which will complicate the mapping process. Note, furthermore, that only the planet's permanent day-side could be mapped using reflected light, in such a case. To constrain the latitude-dependence of albedo, one would need a high-obliquity planet and observations spanning many different phases. For our analysis we use two classes of model maps: one constructed from sinusoidal variations in albedo as a function of longitude, and the other with uniform longitudinal slices of constant albedo \citep{Cowan_2008}. In both cases the albedo is constant with latitude. We compare these two models explicitly in Appendix~II. 

The transformation from to $A(\phi)$ to $A^{*}(t)$ is essentially a low-pass filter which preferentially preserves information about large-scale color variations on the planet. The smoothing kernel (the product of $V$ and $I$) has a FWHM of 76$^{\circ}$ for the March observation and 67$^{\circ}$ for the June observation.  Furthermore, the transformation can be particularly insensitive to certain modes and these cannot be recovered from the light curve inversion \citep[for example, a planet observed at full phase has invisible odd sinusoidal modes,][]{Russell_1906, Cowan_2008}. Even in the idealized toy example discussed in Appendix~II, therefore, the deconvolution can only recover the broadest trends.

\begin{figure}[htb]
\includegraphics[width=84mm]{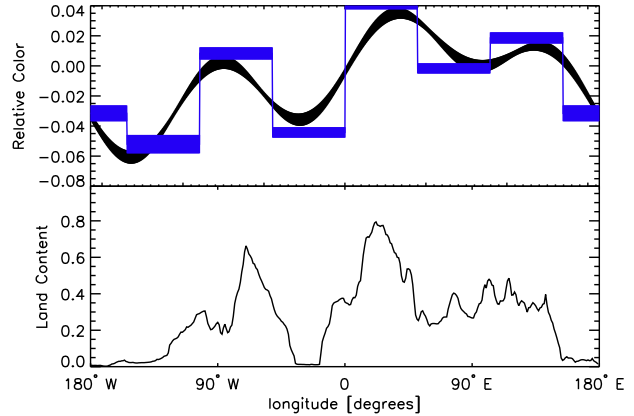}
\caption{Comparison of N-slice (blue) and sinusoidal (black) longitudinal maps of the red eigencolor based on the March EPOXI observations (top panel) to the MODIS cloud-free, equator-weighted distribution of continents (bottom panel). The thickness of the lines in the top panel show the $1\sigma$ uncertainty on the maps. The red eigencolor is particularly sensitive to arid regions (the South-West USA, the Sahara and Middle East, and Australia) since they are consistently cloud-free.}
\label{7_100_Earth1.land_wMODIS_map}
\end{figure}

\begin{figure}[htb]
\includegraphics[width=84mm]{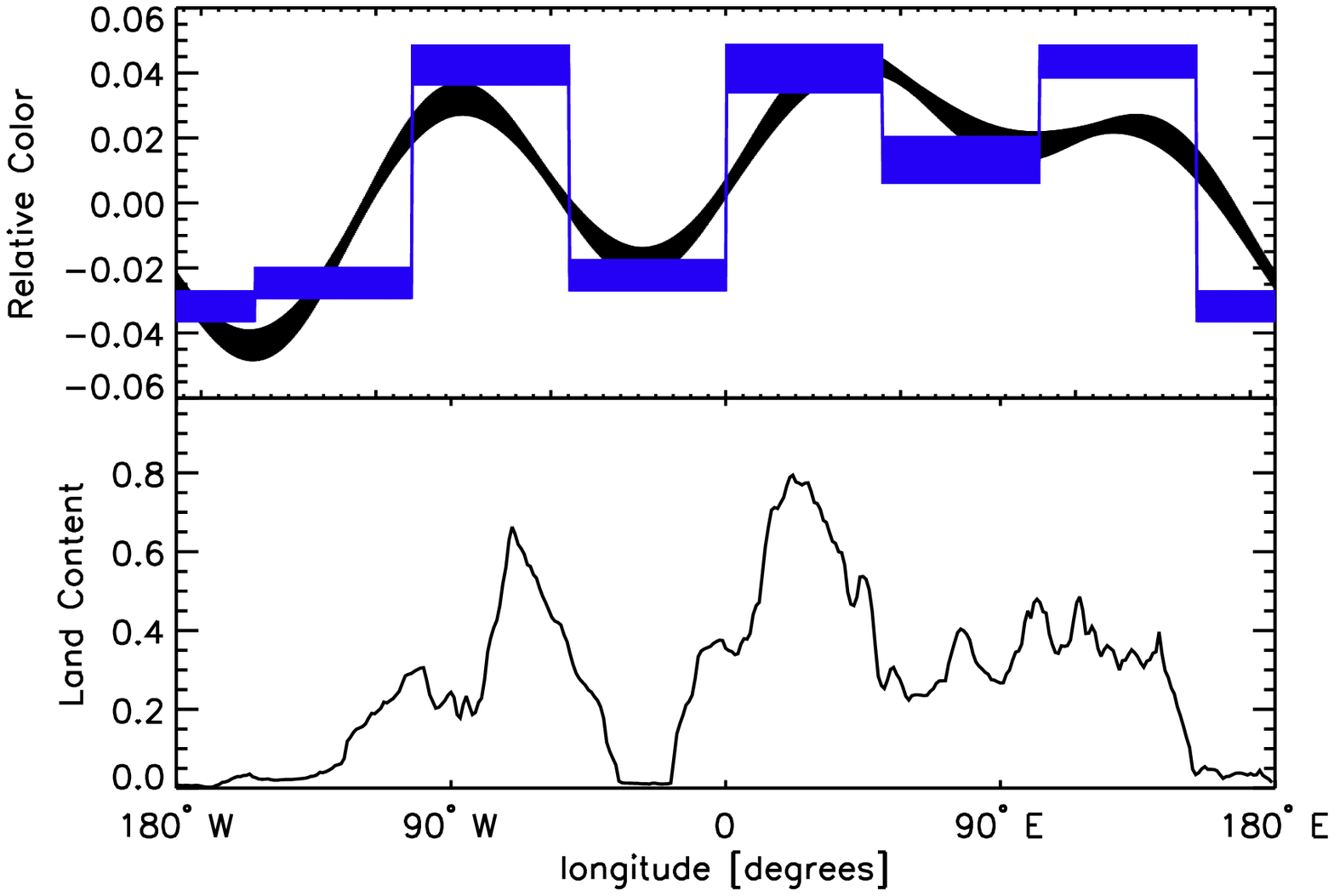}
\caption{Comparison of N-slice (blue) and sinusoidal (black) longitudinal maps of the red eigencolor based on the June EPOXI observations (top panel) to the MODIS cloud-free, equator-weighted distribution of continents (bottom panel). The thickness of the lines in the top panel show the $1\sigma$ uncertainty on the maps. The red eigencolor is particularly sensitive to arid regions (the South-West USA, the Sahara and Middle East, and Australia) since they are consistently cloud-free.}
\label{7_100_Earth5.land_wMODIS_map}
\end{figure}

\begin{figure}[htb]
\includegraphics[width=84mm]{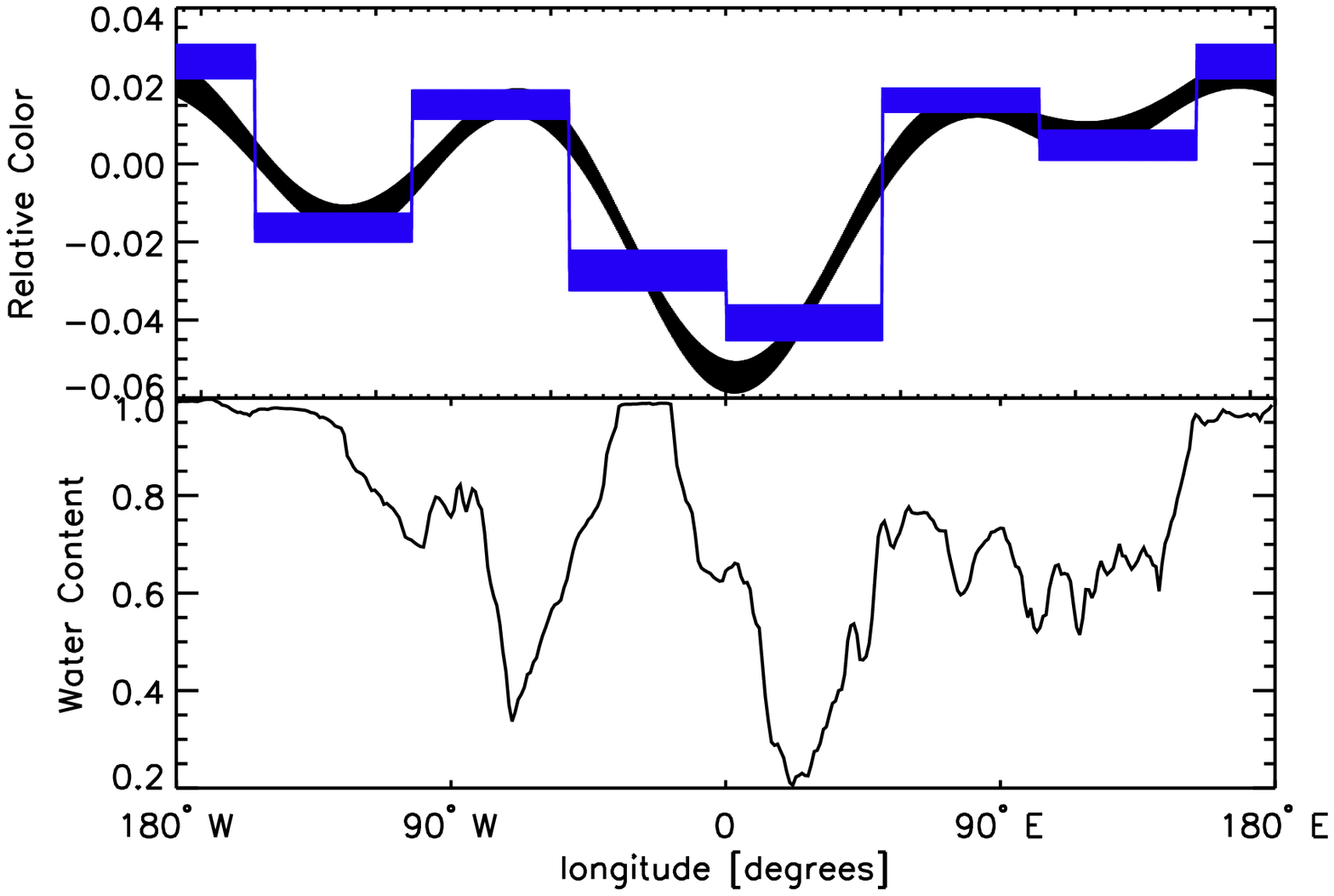}
\caption{Comparison of N-slice (blue) and sinusoidal (black) longitudinal maps of the blue eigencolor based on the March EPOXI observations (top panel) to the MODIS cloud-free, equator-weighted distribution of oceans (bottom panel). The thickness of the lines in the top panel show the $1\sigma$ uncertainty on the maps. The fit to Earth's latitudinaly averaged water content is not excellent, notably the Atlantic Ocean and the Americas. The discrepancy is due to cloud cover, as discussed in the text.}
\label{7_100_Earth1.water_wMODIS_map}
\end{figure}

\begin{figure}[htb]
\includegraphics[width=84mm]{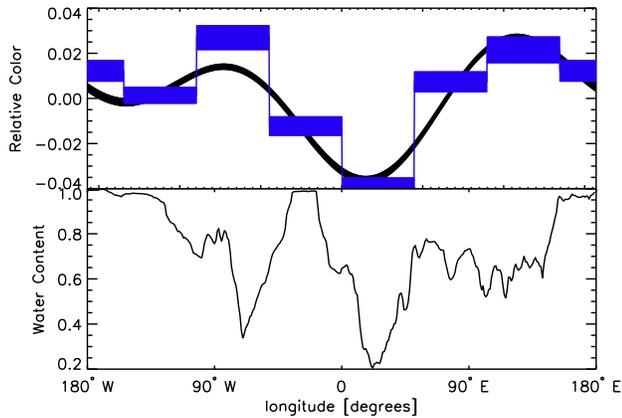}
\caption{Comparison of N-slice (blue) and sinusoidal (black) longitudinal maps of the blue eigencolor based on the June EPOXI observations (top panel) to the MODIS cloud-free, equator-weighted distribution of oceans (bottom panel). The thickness of the lines in the top panel show the $1\sigma$ uncertainty on the maps. The fit to Earth's latitudinaly averaged water content is not excellent, notably the Atlantic Ocean and the Americas. The discrepancy is due to cloud cover, as discussed in the text.}
\label{7_100_Earth5.water_wMODIS_map}
\end{figure}

These best-fit parameters and their uncertainties are determined using a Markov Chain Monte Carlo (MCMC), optimized to have acceptance probabilities near 25\% \citep[e.g.,][]{Ford_2005}. The uncertainty in the sinusoidal model coefficients tends to increase with the harmonic index $n$ because high-frequency modes have a relatively small impact on the observed light curves. We truncate the series when we have sufficient terms to get a reduced $\chi^{2}$ of order unity. In this study we find that including modes up to $n=3$ or $4$ is sufficient, depending on the phase angle and the color being used. By the same token, 7-slice and 9-slice models are used in our fits since they have the same number of free parameters. The sinusoidal and N-slice longitudinal maps of the red and blue eigencolors (roughly, land coverage and ocean coverage) we construct using PCA and light curve inversion are shown in Figures~\ref{7_100_Earth1.land_wMODIS_map}--\ref{7_100_Earth5.water_wMODIS_map}. The March and June maps differ due to different cloud cover on the two days. Nevertheless, the broad peaks and troughs occur at the same longitudes at both epochs, indicating that the eigencolors are sensitive to permanent surface features, not merely clouds. The red eigencolor is more sensitive than the blue eigencolor to the positions of continents and oceans on Earth, despite the fact that both eigencolors can be fooled by clouds. This is because the red eigencolor is most sensitive to the near-infrared light that arid regions of Earth reflect at, and those regions are generally cloud-free.

\begin{figure}[htb]
\includegraphics[width=84mm]{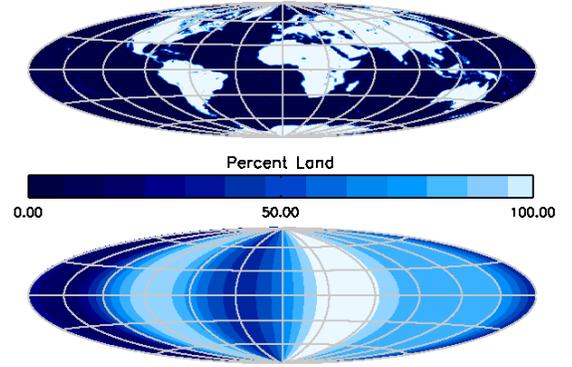}
\caption{Aitoff projection showing the land distribution on Earth in a cloud-free MODIS map (top panel) and the distribution of land as determined from the June disc-integrated EPOXI light curves (bottom panel). The EPOXI map has a longitudinal resolution of approximately $60^{\circ}$; it has no latitudinal resolution, but is weighted toward the equator due to viewing geometry.}
\label{combined_surface_maps}
\end{figure}

\begin{figure}[htb]
\includegraphics[width=84mm]{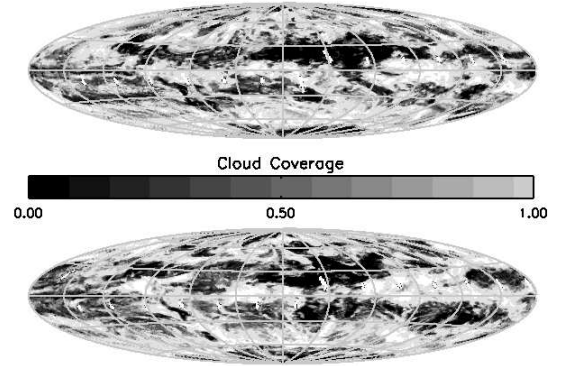}
\caption{Aitoff projection showing the fractional cloud coverage for 03/18/2008 (top panel) and 06/04/2008 (bottom panel), based on MODIS Aqua. The regularly spaced white artifacts represent missing data.}
\label{cloud_maps}
\end{figure}

For our baseline model we assume that the planet's rotation axis is perpendicular to its orbital plane (zero obliquity) and determine the best-fit longitudinal map in each of the two principal colors, assuming the same underlying map for the March and June observations. The June map of the red eigencolor, the most important of the principal components, is shown in Figure~\ref{combined_surface_maps} compared to a cloud-free MODIS\footnote{http://modis.gsfc.nasa.gov/} map of land coverage. Cloud cover, shown in Figure~\ref{cloud_maps}, keeps the match from being perfect, but our blind analysis of the light curves clearly picks out the Atlantic and Pacific Oceans, as well as the major landforms: the Americas, Africa, and Asia.

\subsection{Obliquity}
The obliquity of exoplanets, the angle between their rotational and orbital axes, will not be known \emph{a priori}. The simplest assumption ---and the one used in our baseline model--- is zero-obliquity, but this cannot be assumed as a rule. We fit the light curves using different assumed orbital axes and verify that the longitudinal maps are insensitive to this assumption. For example, the longitudinal maps obtained assuming the Earth's correct obliquity of $23.5^{\circ}$ are almost indistinguishable from our baseline (no obliquity) case. This is not surprising. The apparent albedo of the planet amounts to a weighted average albedo of the regions of the disc that are both visible and illuminated. During a single rotational period all longitudes of the planet will be both visible and illuminated for some fraction of the time. The albedo variations on an oblique planet must be larger to account for the same observed light curves, however. We see this effect in our models: the longitudinal maps for the $23.5^{\circ}$ obliquity  have slightly greater amplitude variations than those for the zero obliquity case, while in the case of $90^{\circ}$ obliquity the subtle changes in apparent albedo in Figure~\ref{combined_lightcurves} can only be explained by enormous (100\%) changes in albedo from one region of the planet to another. 

\begin{figure}[htb]
\includegraphics[width=84mm]{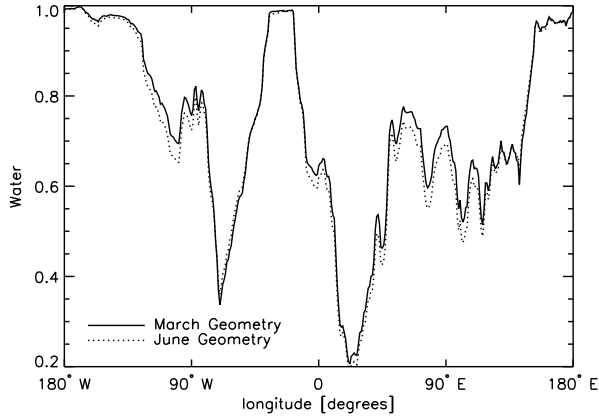}
\caption{The effect of a change in viewing geometry: the equator-weighted longitudinal water distribution on Earth based on the MODIS map is shown in the solid line. The dotted line shows the same but weighted in favor of $11^{\circ}$~N, appropriate for a viewer above the equator near summer solstice. The effect is effectively negligible, justifying our assumption of zero obliquity.}
\label{MODIS_obliquity}
\end{figure}

Obliquity also determines which regions of the planet can be mapped. As long as the planet's rotation axis resides in the sky plane, the reflected light we observe is preferentially coming from near the equator. If instead the planet's rotational axis is not in the sky plane, we preferentially ``see'' some non-zero latitude. A longitudinal map can only be a good representation of the latitudes that are both visible and illuminated, and the latter will change depending on the phase of the planet. During the March EPOXI observations, the sub-solar and sub-observer points were both close to the equator; for the June observations the sub-observer was again nearly equatorial but the sub-solar point was at $22^{\circ}$~N of the equator (it being northern summer). The minuscule effect of this change in viewing geometry can be seen in Figure~\ref{MODIS_obliquity}. The differences between the March and June observations must therefore be due to different cloud cover. 

\section{Discussion}
Spectra of habitable terrestrial exoplanets will tell us the structure and composition of their atmosphere, but may require integration times of weeks to months. Since this is longer than the rotation period of most planets, spectroscopy can only tell us about the spatially-averaged planet. Photometric light curves, with integration times of hours to days, have the potential to reveal spatial variations in the planet's properties. Simulations by \cite{Ford_2001} indicated that diurnal variability in the albedo of an unresolved, cloudless exoplanet could be used to determine its ocean versus land fraction. But in their models it was the specular nature of oceans, rather than their blue color, that distinguished them from continents. The more detailed work of \cite{Williams_2008} indicates that on a cloudy world like Earth, the contribution of specular reflection from oceans will be tiny compared to the diffuse reflection from clouds. The models of \cite{Palle_2008} showed that despite changes in cloud cover, the diurnal albedo variations of an Earth-like planet could be used to determine it's rotation rate. Our study shows that with observations qualitatively similar to those considered by \cite{Palle_2008}, but with greater signal-to-noise and better spectral resolution, it is possible to actually map the longitudinal distribution of colors ---and by extension the dominant surface types--- of Earth.

Earth clouds have higher albedo at all seven wavebands than ocean or continents (see first the bottom panel of Figure~\ref{eigenspectra}), and most regions of the planet have variable cloud cover \citep{Palle_2008}. Observations of multiple consecutive planetary rotations would yield multiple similar longitudinal maps. If differences between these maps were attributed to changes in cloud cover it would be possible to create maps of average cloud cover at each longitude, as well as maps of cloud variability. It may even be possible to partially ``remove'' clouds since the lowest albedo at each longitude would correspond to the observation with the least cloud cover. It would be impossible, however, to strip the clouds from regions that are permanently shrouded (e.g.: tropical rain forests). Insofar as such clouds are permanent features, they can be mapped as terrain features, like oceans and continents.

Clouds are doubly important because they change over time and dominate the total albedo of Earth. However, insofar as roughly half of Earth is cloud-covered at any point in time, the bolometric albedo of the planet does not change much over 24 hrs (as shown in bottom right panel of Figure 1). What can change more significantly is the reflected \emph{color} of the planet, which is what we have studied in this paper. Since clouds are roughly gray, any color (apart from blue Rayleigh scattering) is coming from cloud-free regions. What we have shown in this paper ---without prior assumptions--- is that these cloud-free regions come in two varieties: blue and red.  It is our contention that these correspond to oceans and continents. Moments when both the red and blue eigencolors are high correspond to moments when a particularly high fraction of the visible, illuminated Earth was cloudy.  It is incorrect to think that the color variability is simply a measure of cloudiness, however. Had that been the case, the PCA would only have found a single significant eigencolor, whereas it found two. Despite the fact that clouds dominate the over-all albedo of Earth, the presence of continents and oceans is still discernible in the disc-integrated colors of the planet.

Note that a blue broadband spectrum does not ---in and of itself--- imply liquid water on the surface of a planet (e.g., Neptune).  The \emph{spatial variability} in the blue color is significant, however. An alternative explanation for spatially inhomogeneous blue colors could be partial cloud coverage: the increased path length in regions with fewer clouds could increase the importance of Rayleigh scattering. But the spotty cloud cover on such  planet might, like time-variable cloud cover, give away the presence of water near its condensation point. Furthermore, the blue patches on such a planet might reveal themselves by their very steep blue spectrum. In conjunction with time-averaged spectra, broadband light curves therefore provide a powerful test for the presence of liquid water on a terrestrial planet, and hence habitability.

\section{Conclusions}
In this paper we have shown that despite simplifying assumptions (edge-on geometry, zero-obliquity, diffuse reflection) and the presence of obscuring clouds, one can use time-resolved photometry to detect and map vast blue surfaces, separated by large red regions. If we saw such features on an extrasolar terrestrial planet it would strongly suggest the presence of continents and oceans, and indicate that the planet was a high priority for spectroscopic follow-up.

Although Earth is most reflective at short wavelengths due to
Rayleigh scattering, the wavelengths longward of 700~nm provide the
most spatial information because the relative diurnal variability at
those wavelengths is greatest \citep[25--30\%, in qualitative agreement with the simulations of][]{Ford_2001}, as shown by the error bars in Figure~\ref{mean_spectrum}. The relative variability is
greater for smaller illuminated fraction, in accordance with geometric
considerations. The logical conclusion is that observations of
exoplanets at crescent phase would provide the greatest spatial
resolving power, but this is not true in practice due to poorer
signal-to-noise ratio: the flux from a crescent planet is smaller than at quadrature and at small angular separations the planet is lost in the glare of its host star. Note that interesting measurements can be made if the planet passes \emph{directly} in front of (transit) or behind (secondary eclipse) its host star, but for habitable terrestrial planets the odds of this are not very good (e.g., an Earth-analog has a $0.5$\% probability of transiting a Sun-like host star).

The unknown obliquity of exoplanets does not represent a serious obstacle to mapping their longitudinal color variations, although it will affect which parts of the planet are being mapped. There are pathological orbital configurations that would prohibit mapping (e.g., pole-on rotation axis observed at full phase) but planets in such configurations will be impossible to observe in any case.  Planets will typically be observed near quadrature, so longitudinal maps will not depend sensitively on obliquity.   

\acknowledgments
N.B.C. is supported by the Natural Sciences and Engineering Research Council of Canada. E.A. is supported by National Science Foundation CAREER Grant No. 0645416. This work was supported by the NASA Discovery Program. N.B.C. acknowledges useful discussions of PCA with A.J.~Connolly and J.~VanderPlas, discussions of rotation matrices with H.~Haggard, help from D.~Fabrycky in pointing out an old reference and from P.~Kundurthy in estimating coronagraph S/N.

\section*{Appendix I: Visibility \& Illumination}
One can define a right-handed orthonormal coordinate system, $\hat{x}$, $\hat{y}$, $\hat{z}$, in the observer's inertial reference frame with the origin at the center of the planet, the $x$-axis extending towards the observer, and the $y$- and $z$-axes in the sky plane. The orbital and rotational angular velocity vectors of the planet in this frame are $\vec{\omega}_{\rm orb}$ and $\vec{\omega}_{\rm rot}$, respectively.  If the planet is in an eccentric orbit, the amplitude of $\vec{\omega}_{\rm orb}$ will be a function of time, but in any case the direction of the vector is constant. In the interest of simplicity we consider a circular orbit. Neglecting precession, $\vec{\omega_{\rm rot}}$ will be a constant vector (note that this is the rotation of the planet \emph{in an inertial frame}, not the rotation with respect to the host star). We define the position of the sub-stellar point at $t=0$ as $\hat{x}_{\rm star}$ and the intersection of the prime meridian and the equator on the planet as $\hat{x}_{\rm equ}$. Note that  $\hat{x}_{\rm star}$ is always perpendicular to $\vec{\omega}_{\rm orb}$, and $\hat{x}_{\rm equ}$ is always perpendicular to $\vec{\omega}_{\rm rot}$.

At $t=0$ we define the orthonormal coordinate system fixed with respect to the planet surface $\hat{u}_{0} = \hat{x}_{\rm equ}$, $\hat{w}_{0} = \hat{\omega}_{\rm rot}$ and $\hat{v}_{0} = \hat{w}_{0} \times \hat{u}_{0}$. Likewise, we define at $t=0$ the orthonormal vectors defining the star reference frame: $\hat{d}_{0} = \hat{x}_{\rm star}$, $\hat{f}_{0} = \hat{\omega}_{\rm orb}$ and $\hat{e}_{0} = \hat{f}_{0} \times \hat{d}_{0}$.

The position of a region on the planet can be described by its complementary latitude, $\theta$, measured from the planet's north pole, and its east longitude, $\phi$, measured along the equator from the prime meridian. This patch has a position in the observer frame $\hat{r} = \sin\theta\cos\phi \hat{u} + \sin\theta\sin\phi \hat{v} + \cos\theta \hat{w}$. The visibility of this region from the perspective of the observer is $V(\theta, \phi, t) = \max[\hat{r}\cdot\hat{x}, 0]$.  The inner product can be expanded as:
\begin{equation}
\begin{array}{ll} \hat{r} \cdot \hat{x} = & \sin\theta\cos(\phi + \omega_{\rm rot}t) \hat{u}_{0} \cdot \hat{x}\\
& + \sin\theta\sin(\phi + \omega_{\rm rot}t) \hat{v}_{0} \cdot \hat{x}\\
& + \cos\theta \hat{w}_{0} \cdot \hat{x}.
\end{array}
\end{equation}

The illumination of this patch is $I(\theta, \phi, t) = \max[\hat{r} \cdot \hat{d}, 0]$. The inner product can be expanded as:
\begin{equation}
\begin{array}{ll} \hat{r} \cdot \hat{d} = & \sin\theta\cos(\omega_{\rm orb}t)\cos(\phi + \omega_{\rm rot}t) \hat{u}_{0} \cdot \hat{d}_{0}\\
& + \sin\theta\cos(\omega_{\rm orb}t)\sin(\phi + \omega_{\rm rot}t) \hat{v}_{0} \cdot \hat{d}_{0}\\
& + \sin\theta\sin(\omega_{\rm orb}t)\cos(\phi + \omega_{\rm rot}t) \hat{u}_{0} \cdot \hat{e}_{0}\\
& + \sin\theta\sin(\omega_{\rm orb}t)\sin(\phi + \omega_{\rm rot}t) \hat{v}_{0} \cdot \hat{e}_{0}\\
& + \cos\theta\cos(\omega_{\rm orb}t) \hat{w}_{0} \cdot \hat{d}_{0}\\
& + \cos\theta\sin(\omega_{\rm orb}t) \hat{w}_{0} \cdot \hat{e}_{0}.
\end{array}
\end{equation}

The visibility and illumination can be expressed compactly in terms of the sub-observer longitude, $\phi_{\rm obs}(t) = \phi_{\rm obs}(0) - \omega_{\rm rot} t$, and the constant sub-observer latitude, $\theta_{\rm obs}$, as well as the sub-stellar longitude, $\phi_{\rm star}$, and latitude, $\theta_{\rm star}$:
\begin{equation}
\begin{array}{ll} V = &\max[\sin\theta \sin\theta_{\rm obs} \cos(\phi - \phi_{\rm obs}) + \cos\theta \cos\theta_{\rm obs}, 0]\\
I = & \max[\sin\theta \sin\theta_{\rm star} \cos(\phi-\phi_{\rm star}) + \cos\theta \cos\theta_{\rm star}, 0],\\
\end{array}
\end{equation}
where the sub-stellar longitude is related to the orbital phase measured from the solstice, $\xi(t) = \xi_{0} + \omega_{\rm orb} t$, and the constant planetary obliquity, $\theta_{\rm obl}$, by $\cos\theta_{\rm star} = \cos\xi\sin\theta_{\rm obl}$.

\section*{Appendix II: Comparing N-Slice and Sinusoidal Maps}
\begin{figure}[htb]
\includegraphics[width=84mm]{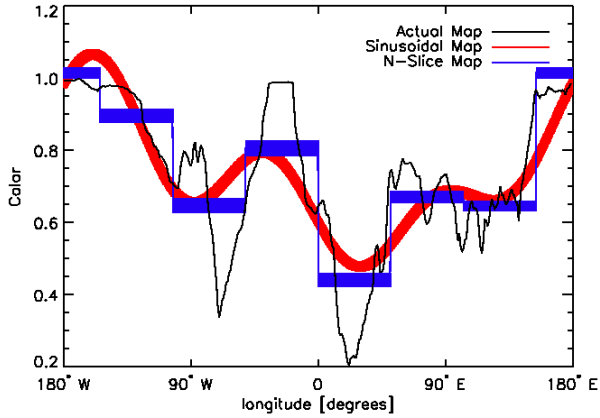}
\caption{Comparison of sinusoidal and N-slice longitudinal maps. The thickness of the lines denotes the $\pm 1\sigma$ intervals. The basic features of the underlying map (the major continents and oceans) are recovered by either mapping technique.}
\label{slice_sine}
\end{figure}

\begin{figure}[htb]
\includegraphics[width=84mm]{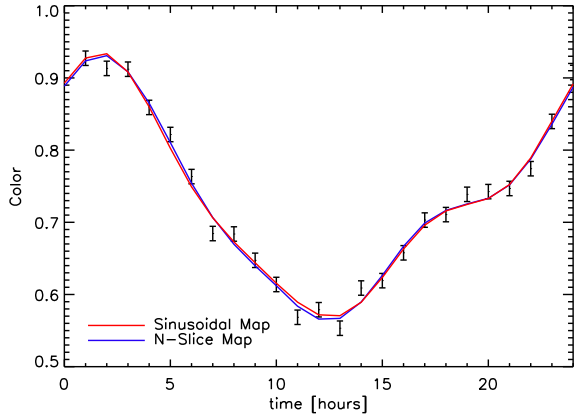}
\caption{Comparison of the light curves produced by the three maps of Figure~\ref{slice_sine}.}
\label{slice_sine_lc}
\end{figure}

The N-Slice and sinusoidal models are compared in Figure~\ref{slice_sine}, with the resulting light curves shown in Figure~\ref{slice_sine_lc} \citep[see also][]{Cowan_2008}. A MODIS map of liquid water content (s.f. top panel of Figure~\ref{combined_surface_maps}) was integrated to a one-dimensional, equator-weighted map of water content, the black line in Figure~\ref{slice_sine}. A model light curve, shown in black on Figure~\ref{slice_sine_lc}, was generated assuming photometric uncertainties of 1\%. This light curve became the input for light curve inversions using sinusoidal (red) and N-slice (blue) maps, shown here with $\pm 1\sigma$ intervals. 

\section*{Appendix III: Planet Mapping in the Specular Regime}\label{specular}
An interesting limiting case arises when the entirety of the light from the planet originates from the glint spot where the product $VI$ is maximized, as would be the case for purely specular reflection. The latitude of the specular point is given by 
\begin{equation}
\cos\theta_{\rm spec} = \frac{\cos\theta_{\rm star} + \cos\theta_{\rm obs}}{\sqrt{2[1+\sin\theta_{\rm star} \sin\theta_{\rm obs}\cos(\phi_{\rm star}-\phi_{\rm obs}) + \cos\theta_{\rm star}\cos\theta_{\rm obs}]}}
\end{equation}
and its longitude is given by
\begin{equation}
\tan\phi_{\rm spec} = \frac{\sin\phi_{\rm star}\sin\theta_{\rm star} + \sin\phi_{\rm obs}\sin\theta_{\rm obs}}{\cos\phi_{\rm star}\sin\theta_{\rm star} + \cos\phi_{\rm obs}\sin\theta_{\rm obs}}.
\end{equation}

\begin{figure}[htb]
\includegraphics[width=84mm]{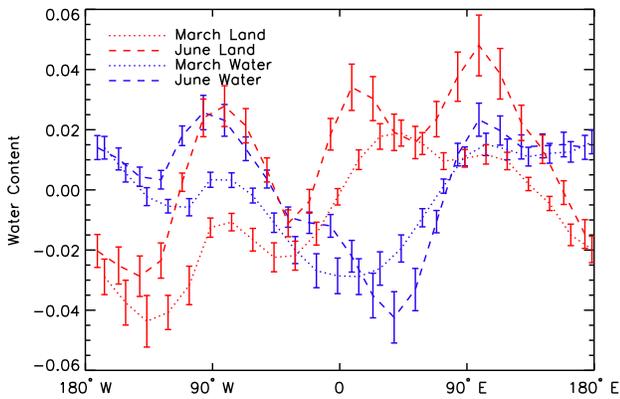}
\caption{Longitudinal maps of land (red eigencolor) and water (blue eigencolor) on Earth, based on a principal component analysis of disc-integrated light curves and the assumption of specular reflection.}
\label{simple_map}
\end{figure}

In Figure~\ref{simple_map} we show the results of specular inversion on the light curves shown in Figure~\ref{eigenprojections}. We have assumed edge-on, zero-obliquity geometry. The map of the red eigencolor shows three distinct peaks, corresponding to (from left to right) North America, Africa and Asia. The map of the blue eigencolor is characterized by a high plateau between 90$^{\circ}$ E to 90$^{\circ}$ W, which corresponds to the Pacific Ocean. The diffusely and specularly reflecting cases bracket more realistic scattering phase functions, with Solar System planets and moons being closer to the former. The detectability of the major continents and oceans on Earth using \emph{either} assumption indicates the robustness of the result.

\end{document}